\documentclass{emulateapj}
\usepackage{lscape}

\shorttitle{COHERENT RADIO EMISSION FROM ULTRACOOL DWARFS}
\shortauthors{Hallinan et al.}

\begin{document}


\title{CONFIRMATION OF THE ELECTRON CYCLOTRON MASER INSTABILITY AS THE DOMINANT SOURCE OF RADIO EMISSION FROM VERY LOW MASS STARS AND BROWN DWARFS}


\author{G. Hallinan\altaffilmark{1},  A. Antonova\altaffilmark{2}, J.G. Doyle\altaffilmark{2}, S. Bourke\altaffilmark{1,3}, C. Lane\altaffilmark{1}  and A. Golden\altaffilmark{1}}

\altaffiltext{1}{Centre for Astronomy, National University of Ireland, Galway, Ireland; gregg@it.nuigalway.ie, stephen@it.nuigalway.ie, c.lane2@nuigalway.ie, agolden@it.nuigalway.ie}

\altaffiltext{2}{Armagh Observatory, College Hill, Armagh BT61 9DG, N. Ireland; tan@arm.ac.uk, jgd@arm.ac.uk}

\altaffiltext{3}{Joint Institute for VLBI in Europe, Postbus 2, 7990 AA Dwingeloo, the Netherlands; stephen@it.nuigalway.ie}




\begin{abstract}
We report on radio observations of the M8.5 dwarf LSR J1835+3259 and the L3.5 dwarf 2MASS J00361617+1821104, which provide the strongest evidence to date that the electron cyclotron maser instability is the dominant mechanism producing radio emission in the magnetospheres of ultracool dwarfs. As has previously been reported for the M9 dwarf TVLM 513-46546, periodic pulses of $100\%$ circularly polarized, coherent radio emission are detected from both dwarfs with periods of $2.84 \pm 0.01$ and $3.08 \pm 0.05$ hours respectively for LSR J1835+3259 and 2MASS J00361617+1821104. Importantly, periodic \textit{unpolarized} radio emission is also detected from 2MASS J00361617+1821104, and brightness temperature limitations rule out gyrosynchrotron radiation as a source of this radio emission. The unpolarized emission from this and other ultracool dwarfs is also attributed to electron cyclotron maser emission, which has become depolarized on traversing the ultracool dwarf magnetosphere, possibly due to propagations effects such as scattering. Based on available $v \sin i$ data in the literature and rotation periods derived from the periodic radio data for the three confirmed sources of electron cyclotron maser emission, TVLM 513-46546, LSR J1835+3259 and 2MASS J00361617+1821104, we determine that the rotation axes of all three dwarfs are close to perpendicular to our line of sight. This suggests a possible geometrical selection effect due to the inherent directivity of electron cyclotron maser emission, that may account for the previously reported relationship between radio activity and $v \sin i$ observed for ultracool dwarfs. We also determine the radius of the dwarf LSR J1835+3259 to be $\geq 0.117 \pm 0.012$ R$_{\odot}$. The implied size of the radius, together with the bolometric luminosity of the dwarf, suggests that either LSR J1835 is a young or intermediate age brown dwarf or that current theoretical models underestimate the radii of ultracool dwarfs.

\end{abstract}





\section{INTRODUCTION}

\par In recent years, the techniques used in the study of magnetic activity in dwarf stars have been adapted and applied to ultracool dwarfs (late M, L, and T dwarfs). Surveys of chromospheric H$\alpha$ and coronal X-ray emission have revealed a sharp decline in $L_{H\alpha}/L_{bol}$ and $L_{X}/L_{bol}$ beyond spectral type M7, such that evolved L and T dwarfs have relative luminosities two orders of magnitude lower than observed for saturated activity in earlier type M dwarfs \citep{neuhauser99, fleming03, gizis00, mohanty03, stelzer04, west04, stelzer06a, schmidt07}. This reduction in activity may be associated with the decreasing fractional ionization of the photosphere and atmosphere with later spectral type, such that magnetic fields becomes decoupled from the increasingly cool and neutral atmosphere, reducing the dissipation of magnetic field energy and hence the heating of plasma \citep{meyer99, mohanty02}.

\par Importantly, this reduction in the non-radiative heating of plasma to chromospheric and coronal temperatures does not necessarily imply a drop in magnetic field strengths and filling factors. Indeed, many lines of evidence suggest the contrary, confirming that efficient magnetic dynamo action is sustained for at least some fraction of the ultracool dwarf population. Although luminosities are much lower than for earlier type M dwarfs, quiescent H$\alpha$ emission is still detectable from a number of L and T dwarfs, and \citet{audard07} have recently confirmed the first detection of quiescent X-ray emission from an L dwarf, the binary system, Kelu-1. Extremely energetic flares have also been detected in optical, UV, and X-ray bands as well as in spectral line emission such as H$\alpha$, providing strong evidence for magnetic reconnection events \citep{reid99, burgasser00, gizis00, rutledge00, liebert03, fuhrmeister04, stelzer04, rockenfeller06, stelzer06b, schmidt07}. 

\par Perhaps, the strongest indicator of magnetic activity in ultracool dwarfs, has been the detection of both quiescent and flaring non-thermal radio emission from a number of late M and L type dwarfs \citep{berger01, berger02, berger05, berger06, berger07a, berger07b, burgasser05, osten06, phan-bao07, hallinan06, hallinan07, antonova07}. The radio luminosities of ultracool dwarfs are of the same order as those inferred for earlier type, active M dwarf stars, with no indication of a sharp drop in emission at spectral type M7, such as that observed in H${\alpha}$ and X-ray emission. Therefore, the tight correlation observed between the X-ray and centimetric radio luminosities of active late-type dwarf stars, such that $L_X/L_R \approx 10^{15\pm 1}$ Hz \citep{gudel93, benz94}, breaks down for spectral type $\gtrsim$ M7. 

\par Thus far, 9 dwarfs in the spectral range M7-L3.5, including one binary system, have been detected in surveys encompassing $\sim$ 100 targets. Two of these 9 ultracool dwarfs, the M9 dwarf TVLM 513-46546 (hereafter TVLM 513) and the L3.5 dwarf 2MASS J00361617+1821104 (hereafter 2MASS J0036), have been particularly well studied in further deep pointings and multifrequency observations that have revealed much on the nature of the radio emission \citep{berger05, berger07a, osten06, hallinan06, hallinan07}. Both dwarfs have been found to be sources of broadband radio emission, a component of which is found to be highly circularly polarized and periodically variable, with periods of $\sim 3$ and $1.96$ hours for 2MASS J0036 and TVLM 513 respectively \citep{berger05, hallinan06, hallinan07}. These periodicities were found to be consistent with the putative rotation periods of the dwarfs derived from $v \sin i$ data in the literature, which would place strong constraints on the brightness temperature and directivity of the associated emission mechanism \citep{hallinan06}. In order to account for this high brightness temperature, implicit high directivity and high degree of circular polarization, \citet{hallinan06} suggested that the periodic, highly circularly polarized component of the radio emission may be produced at the polar regions of a large scale magnetic field, by a coherent process, the electron cyclotron maser (hereafter ECM) instability \citep[][and references therein]{treumann06}, the same mechanism known to be responsible for the radio emission at kHz and MHz frequencies from the magnetized planets in our solar system \citep{ergun00, zarka98} and also thought to be a source of certain classes of solar and stellar bursts \citep{melrose82, bingham01, kellett02}. The resulting emission is intrinsically highly circularly polarized and strongly beamed at large angles to the magnetic field in the source region which, together with rotation of the dwarf, accounts for the observed periodicity. 

\par Follow-up observations of TVLM 513 found the dwarf in a much more active state, with an order of magnitude increase observed in the flux density of the periodic, highly circularly polarized component of the radio emission detected from the dwarf \citep{hallinan07}. Most notably, this was manifested in the detection of multiple periodic, narrow duty cycle pulses of $100\%$ circularly polarized emission, the brightest of which had a mean flux density of $\sim 5$ mJy. On this occasion, simultaneous photometric monitoring observations confirmed that the periodicity was indeed associated with the period of rotation of the dwarf \citep{lane07}. The pulses, which were present over the duration of the 10 hour observation, were conclusively coherent in nature and found to be consistent with generation by the ECM instability, requiring the ultracool dwarf to possess magnetic fields of strength $\geq 3$ kG, as high as those possessed by earlier type classical dMe flare stars. It was recently confirmed that ultracool dwarfs do indeed possess such high strength magnetic fields by \citet{reiners07}, who studied the magnetically sensitive Wing-Ford FeH band in a wide spectral range of M dwarfs, thus extending the direct measurement of magnetic field strengths out to M9. Although this study did not encompass L or T dwarf candidates, the results confirm that integrated surface magnetic flux is at least as high in late M dwarfs, such as TVLM 513, as in active earlier type M dwarfs.

\par The narrow bunching in phase of multiple pulses of both left and right $100\%$ circularly polarized radio emission detected from TVLM 513, which originate in regions of opposite magnetic polarity, have also revealed the likely presence of a dipolar component to the large-scale magnetic field \citep{hallinan07}. Zeeman Doppler Imaging (ZDI) has previously detected such a large-scale dipolar field structure, stable over timescales $> 1$ year, for the rapidly rotating, fully convective M4 dwarf star V374 Peg \citep{donati06, morin07}. Further studies of other M dwarfs in the spectral range M4-M7 have confirmed that strong axisymmetric large-scale field topologies are indeed associated with rapid rotation in fully convective objects \citep{donati07}. Although, the ZDI technique has not, as yet, been specifically applied to late M or L dwarfs, it is noteworthy that TVLM 513, with a $v \sin i$ of $60 $ km s$^{-1}$, is also a fully convective rapid rotator and should possess a similar type dynamo to V374 Peg.

\par Therefore, ECM emission from the polar regions of a high strength, large-scale magnetic field can account for the periodic detection of highly circularly polarized radio emission from TVLM 513 and is also consistent with both the measured magnetic field strengths for ultracool dwarfs and the observed large-scale fields reported for rapidly rotating fully convective M dwarfs. We also note that this mechanism has previously been invoked for a $100\%$ coherent flare detected from the M9 dwarf, DENIS-P J104814.9-395604 \citep{burgasser05}. However, some important questions remain unanswered. It remains unclear what fraction of ultracool dwarfs are detectable at radio frequencies and what are the physical characteristics distinguishing `active' and `inactive' ultracool dwarfs. Although radio surveys have yielded the detection of 9 active dwarfs in observations of ~100 targets, such surveys have predominantly consisted of short observations of targets at a single frequency of 8.5 GHz, important when considering radio spectra of ultracool dwarfs are not fully characterized, and a number of dwarfs have exhibited a high degree of long term variability. For example, \citet{antonova07} have recently shown that the quiescent radio luminosity of one ultracool can vary by a factor of 10 in 3 separate observations spanning 6 weeks. It is also unclear if the ECM instability is the sole mechanism producing radio emission in ultracool dwarf magnetospheres. All dwarfs detected thus far, including those known to periodically produce highly circularly polarized radio emission such as TVLM 513 and 2MASS J0036, are found to produce quiescent radio emission with low or moderate circular polarization. Therefore, either a component of the ECM emission becomes depolarized on traversing the ultracool dwarf magnetosphere, or a secondary component produced by an incoherent process, such as gyrosynchrotron or synchrotron radiation, is applicable.

\par An important step in determining if ECM emission is indeed the dominant source of radio emission from ultracool dwarfs is increasing the sample of targets selected for extended monitoring observations, with a view to investigating the presence of periodic highly circularly polarized radio emission and/or bright pulses, indicative of ECM emission from a large-scale magnetic field. We have selected two ultracool dwarfs, the M8.5 dwarf LSR J1835+3259 (hereafter LSR J1835) and the L3.5 dwarf 2MASS J0036, for deep pointings with the Very Large Array (VLA) \footnote[1]{The VLA is operated by the National Radio Astronomy Observatory, a facility of the National Science Foundation operated under cooperative agreement by Associated Universities, Inc.}. 

\par LSR J1835 was chosen for a single 11 hour observation at 8.44 GHz, because it is of similar spectral type to the archetypal pulsing ultracool dwarf, TVLM 513, and also due to its close proximity ($< 6$ pc) \citep{reid03}, the intensity of its radio emission in a previous short duration survey observation of $\sim 2$ hours ($0.525 \pm 0.015$ mJy) \citep{berger06} and its favorable declination for a lengthy single observation with the VLA. Furthermore, like TVLM 513, this dwarf has previously showed evidence for variability in I band photometric data \citep{reid03}. In the case of TVLM 513 this variability was later shown to be periodic with rotation of the dwarf, and associated with the presence of magnetic spots, consistent with the high strength magnetic fields required for the detection of ECM emission \citep{lane07}. LSR J1835 was therefore deemed a possible analog of TVLM 513 and a suitable target for a radio monitoring observation. 

\par 2MASS J0036, on the other hand, has previously been observed by \citet{berger05} in multi-epoch observations that showed this target to be a source of periodic highly circularly polarized emission. The authors attributed this periodic emission to gyrosynchrotron radiation from an extended source region larger than the stellar disk with magnetic field strengths in the source region $\sim 175$G. However, \citet{hallinan06, hallinan07} suggested that this periodic emission may also be produced in the same fashion as that detected from TVLM 513, i.e., ECM emission from compact regions at the poles of a large-scale magnetic field with kG field strengths in the source region. We have conducted a 12 hour observation of 2MASS J0036 with the VLA at a frequency of 4.88 GHz in attempt to distinguish between these two mechanisms as the source of the radio emission from this ultracool dwarf.

\section{RADIO OBSERVATIONS}

\par LSR J1835 was observed on 2006 September 18/19 at a frequency of 8.44 GHz for a total of $\sim 11$ hours, including overheads and time spent on the sources 1850+284 and 1331+305 for phase and flux density calibration respectively. 22 dishes of the VLA in B configuration were used in standard continuum mode with 2 $\times$ 50 MHz contiguous bands. Observations of the source were suspended for a 0.9 hour period during which it was above the elevation limit of the VLA. 2MASS J0036 was observed on 2006 September 24 at frequency of 4.88 GHz for a total of $\sim 12$ hours, including overheads and time spent on the sources 0042+233 and 0137+331 for phase and flux density calibration respectively. 25 dishes of the VLA in B configuration were used in standard continuum mode with 2 x 50 MHz contiguous bands. Data reduction was carried out with the Astronomical Image Processing System (AIPS) software package. The visibility data was inspected for quality both before and after the standard calibration procedures, and noisy points were removed. For imaging the data we used the task IMAGR. We also CLEANed the region around each source and used the UVSUB routine to subtract the resulting source models for the background sources from the visibility data. The source was shifted to the phase center and light curves were generated by plotting the real part of the complex visibilities as a function of time. 

\subsection{LSR J1835}

\begin{figure}[ht!]
\plotone{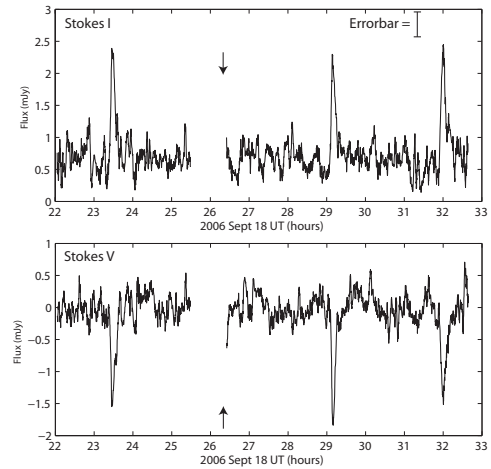}
\caption{The light curves of the total intensity (Stokes I) and the circularly polarized (Stokes V) radio emission detected at 8.44 GHz from LSR J1835+3259 on 2006 Sept 18/19. For the Stokes V light curve, right circular polarization is represented by positive values and left circular polarization is represented by negative values. The data has a time resolution of 10 seconds and is smoothed with a moving window of 150 seconds. A $\sim 0.9$ hour gap is present in the observation during which LSR J1835 was above the elevation limit of the VLA. The source is detected with a mean flux density of $0.722 \pm 0.015$ mJy. As well as a largely non-varying quiescent component of radio emission, three periodic pulses of $100\%$ left circularly polarized emission are detected with $p = 2.84 \pm 0.01$ hours. The arrows highlight the time of a pulse which occurred while the source was above the elevation limit of the VLA. Although the peak of this pulse was missed, the latter part was detected when the observation resumed.}

\end{figure}

\par LSR J1835 was detected as a persistent source over the duration of the 11 hour observation with a flux density of $0.722 \pm 0.015$ mJy (Figure 1). This emission can be resolved into two components, a quiescent, largely unpolarized component and periodic, $100\%$ circularly polarized, coherent pulses that reach a flux density of $\sim 2.5$ mJy. The pulses are periodically present throughout the observation, indicating that their apparent transient nature is not due to intrinsic variability in the source region, but rather due to the rotational modulation of highly beamed coherent emission that is stable over timescales $\geq 11$ hours. The periodicity was established using three independent methods, (1) conducting a Lomb-Scargle periodiogram analysis of both the Stokes I and V light curves (2) phase folding the data over a wide range of periods and then analyzing the $\chi^{2}_r$ values of the resulting light curves in an attempt to detect a peak in variance, and (3) determining the time difference between the peaks of the narrow duty cycle pulses. In total, 3 pulses were detected recurring with a period $p = 2.84 \pm 0.01$ hours. A 4th pulse is also expected to have occurred at September 18 26.3 UT, while the source was above the elevation limit of the VLA. The end of this pulse was indeed detected when observations recommenced at September 18 26.42 UT. \citet{berger07b} have recently reported on a deep pointing of LSR J1835, taken approximately 8 months after the observations reported here, and claim no evidence of periodic or highly circularly polarized emission. Therefore, the degree of activity in the compact regions associated with the pulses can vary greatly over timescales of order a few months, similar to what has previously been confirmed for TVLM 513 \citep{osten06, hallinan06, hallinan07, berger07a}. 

\par Two mechanisms are usually invoked for the generation of solar and stellar coherent radio emission, plasma radiation and the ECM instability. We rule out plasma radiation as a source of the coherent pulses detected from LSR J1835, which is generated at the fundamental and perhaps first harmonic of the plasma frequency, $\nu_p \approx 9000 n_e^{1/2}$ Hz and requires $\nu_p \gg \nu_c$, where $n_e$ is the plasma electron density and $\nu_c$ is the electron cyclotron frequency in the source region \citep{dulk85, bastian98, gudel02}. This mechanism is thought to be a dominant source of radio bursts detected from the Sun, but is generally confined to frequencies below 1-3 GHz as free-free absorption strongly increases with frequency, inhibiting the escape of the propagating emission. The coherent pulses detected from LSR J1835, which were observed at a frequency of 8.44 GHz, are therefore unlikely to be due to plasma radiation. It has been postulated that the optical depth for free-free absorption should be reduced in the coronae of very active stars, such as RS CVn binaries, where thermal coronal temperatures can reach $10{^8}$ K, enabling the escape of plasma radiation at higher frequencies \citep{white95, ostenbastian06}. However, a deep Chandra observation of LSR J1835 did not detect any X-ray emission, precluding the presence of such a high temperature corona \citep{berger07b}. 

\par ECM emission is primarily generated at the electron cyclotron frequency, $\nu_c \approx 2.8 \times 10^6 B$ Hz, and therefore the detection of ECM emission at 8.44 GHz from LSR J1835 requires magnetic fields strengths of 3kG in the source region of the pulses, consistent with the magnetic field strengths previously confirmed for ultracool dwarfs by \citet{hallinan06, hallinan07} and \citet{reiners07}. Contrary to plasma radiation, ECM emission requires source conditions such that $\nu_p \ll \nu_c$. This is particularly notable when considering that X-ray and H$\alpha$ luminosities drop rapidly beyond spectral type M7, indicating progressively cooler and more neutral atmospheres, lower electron plasma densities and hence lower plasma frequencies in the magnetospheres of ultracool dwarfs. Simultaneously magnetic flux, and hence electron cyclotron frequencies, are maintained across the same spectral range. Therefore, magnetospheric conditions where $\nu_p \ll \nu_c$ should become increasingly more prevalent with later spectral type, consistent with the detection of ECM emission from a number of ultracool dwarfs. 

\par The growth rate of ECM emission is angle dependent and predominantly in the X-mode \citep[][and references therein]{treumann06} accounting for the narrow beaming and $100\%$ circular polarization of the pulses detected from LSR J1835. Assuming photospheric equipartition field strengths $\lesssim 10$ kG \citep{chabrier06, browning07} and field divergence in the limiting case of a global dipole with $B(R) \propto R^{-3}$, we can determine that a source with local field strength $\sim 3$ kG is confined to a height $h \leq 0.5$ R above the stellar surface. Studies of planetary ECM have shown the emission to be beamed at a large angle to the local magnetic field, in a conical sheet a few degrees thick \citep{zarka98}. We can therefore define the source size $L \leq 2 \pi (R+h) (d - t/360)$, where $d$ is the duty cycle of the emission and $t$ is the thickness, in degrees, of the wall of the emission cone. In the limiting case of perfect beaming $L \leq 0.57R$, and we can similarly limit the thickness of the wall of the emission cone to be $\leq 22 \arcdeg$. The brightness temperature of the emission is given by 

\begin{equation}T_{\mathrm{B}} = 2\times 10^9(f_{\nu}/mJy)(\nu/GHz)^{-2}(d/pc)^2(L/R_{\mathrm{Jup}})^{-2} K \end{equation}  

Assuming a radius for LSR J1835 of the order $R \approx 0.1 R\odot \sim R_{Jup}$ yields a brightness temperature $\geq 7 \times 10^{9}$ K for perfectly beamed emission. However, we note that this calculation does not take into account the morphology of the source region. In particular, the emission at 8.44 GHz, which is generated primarily at the electron cyclotron frequency, originates from a thin radial slice of the large-scale field of very narrow vertical thickness where the magnetic field $\sim 3$ kG. The pulses detected from LSR J1835 are an order of magnitude lower luminosity than those previously detected from TVLM 513. Although the periodic pulses detected from LSR J1835 are $100\%$ circularly polarized, the average circular polarization over the duration of the 11 hour observation is quite low $\approx 8 \pm 2 \%$. Therefore, as was previously found to be the case for TVLM 513, a component of the radio emission detected from LSR J1835 is largely unpolarized.

\subsection{2MASS J0036}

\begin{figure}[ht!]

\plotone{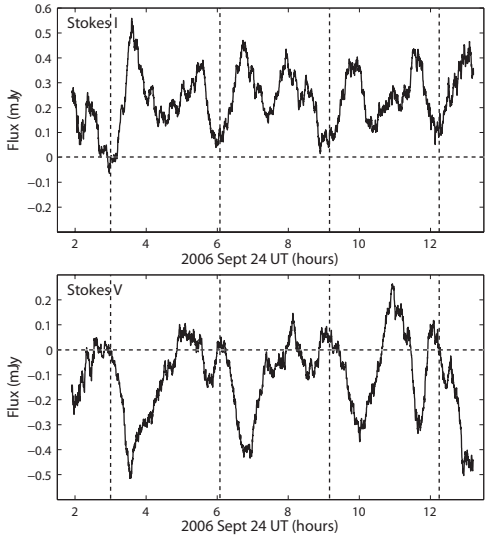}
\caption{The light curves of the total intensity (Stokes I) and the circularly polarized (Stokes V) radio emission detected at 4.88 GHz from 2MASS J00361617+1821104 on 2006 Sept 24. For the Stokes V light curve, right circular polarization is represented by positive values and left circular polarization is represented by negative values. The data has a time resolution of 10 seconds and is smoothed with a moving window of 1000 seconds. The source is detected with a mean flux density of $0.241 \pm 0.014$ mJy and both the Stokes I and V light curves were found to be periodic with $p = 3.08 \pm 0.05$ hours, determined to be the period of rotation of the dwarf. The dashed vertical lines divide the light curve into sections corresponding to this period. The net circular polarization over the duration of the observation is $47 \pm 5 \%$. However, both left and right circularly polarized emission is detected resulting in a total circular polarization exceeding $60\%$.}     
\end{figure}

\par 2MASS J0036 was also detected as a persistent source throughout the 12 hour observation on 2006 September 24 with a flux density of $0.241 \pm 0.014$ mJy (Figure 2). Once again, both the total intensity (Stokes I) and circularly polarized (Stokes V) light curves are found to be periodic with $p = 3.08 \pm 0.05$ hours. Photometric monitoring observations of this ultracool dwarf, conducted a week prior to the radio observations, have yielded I band light curves with a periodicity of $\sim 3$ hours confirming this to be the period of rotation of 2MASS J0036 \citep{lane07}. The degree of circular polarization of the radio emission when averaged over the 12 hour observation is $-47 \pm 5 \%$, with the negative value indicating a net left circular polarization. However, this value does not accurately reflect the amount of polarized radio emission detected from 2MASS J0036 as intervals of both left and right circularly polarized emission are periodically detected that add destructively when summed over the 12 hours, resulting in a lower average circular polarization. From the light curves in Fig. 2 we estimate that as much as $60\%$ of the radio emission is either left or right circularly polarized. Crucially, the degree of circular polarization periodically reaches $100\%$ for large fractions of the period of rotation of the dwarf, indicating that the bulk of the emission, both polarized and unpolarized, is periodic and a quiescent, unvarying component is absent.  

\par The high degree of circular polarization of the radio emission detected from 2MASS J0036 is highlighted in Figure 3 which shows the Stokes I light curve, as shown in Figure 2, directly correlated with the absolute degree of circular polarization, i.e., both left and right circular polarization are represented by positive values. The light curves are overlaid with vertical lines that demarcate intervals corresponding to the period of rotation of the dwarf. The Stokes I light curve is characterized by two broad peaks in emission per period of rotation. The first periodic peak, which we label as the main pulse (MP, Figure 3), is $100\%$  circularly polarized, has a duty cycle $> 30\%$ and is present throughout the observation. Once again, the $100\%$ circular polarization of the emission indicates coherent ECM emission. Assuming emission at the fundamental electron cyclotron frequency requires magnetic field strengths in the source region of $\sim 1.7$ kG, providing the first confirmation of kG magnetic fields in an L dwarf. The main pulse has a much higher duty cycle than the narrow pulses detected from either TVLM 513 or LSR J1835, the latter having duty cycles of a few percent. However, we note that, in the case of TVLM 513, multiple pulses were detected that were confined to a range of phase of rotation of the dwarf of width $\sim 0.35$ \citep{hallinan07}. These pulses were attributed to individual compact source regions confined to the magnetic poles of a large scale dipolar field where the magnetic field strength $\sim 3$kG. In the case of 2MASS J0036, $100\%$ circularly polarized emission is detected continuously over a range of phase of width $\sim 0.3$, indicating a much more extended stable source region, possibly corresponding to a radial slice of the magnetic pole of the dwarf where the magnetic field strength is $\sim 1.7$ kG. We note that the flux density of the main pulse detected from 2MASS J0036 is an order of magnitude lower than the narrow duty cycle pulses detected from TVLM 513 and unresolved substructure may be present.

\begin{figure}[ht!]
\plotone{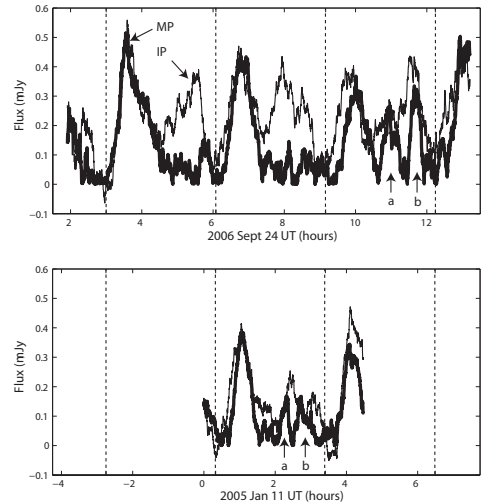}
\caption{[Top] The light solid line represents the total intensity (Stokes I) light curve of the radio emission detected at 4.88 GHz from 2MASS J00361617+1821104 on 2006 Sept 24, as shown in Figure 2. The heavier solid line represents the circularly polarized (Stokes V) light curve as shown in Figure 2. However, in this instance, both right and left circularly polarized emission are represented by positive values to directly highlight the high degree of circular polarization of the radio emission. The dashed vertical lines divide the light curve into sections corresponding to the period of rotation of the dwarf. During each period, the emission is characterized by two broad peaks, which we identify here as the main pulse (MP) and interpulse (IP). Four main pulses and three interpulses are detected over the duration of the 12 hour observation. The main pulse is $100\%$ circularly polarized throughout the observation. The first two detected interpulses are unpolarized but significant polarization is detected in the third. Two sub pulses are resolved in the third interpulse, with the first being right circularly polarized (marked a) and the second being left circularly polarized (marked b). [Bottom] A similar light curve is presented for data obtained 18 months earlier on 2006 Jan 11/12. The $100\%$ circularly polarized main pulse is clearly present as is the interpulse, albeit at lower levels of emission. The interpulse is again characterized by two sub pulses, with the first being right circularly polarized (marked a) and the second being left circularly polarized (marked b). }     
\end{figure}

\par A second broad periodic peak in emission is also detected, which we label as the interpulse (IP, Figure 3). This interpulse is also present over the duration of the observation and is of similar flux density and duty cycle to the main pulse. However, in contrast to the main pulse, the first two interpulses are largely unpolarized. Significant circular polarization is detected in the third interpulse, resolved as two sub pulses in the circularly polarized light curve, the first being right circularly polarized and the second being left circularly polarized, marked as a and b respectively in Figure 3. If we interpret the interpulse emission in the context of incoherent gyrosynchrotron emission produced by a non-thermal population of electrons with a power law energy distribution function, such that $n(E) \propto E^{-\delta}$, the brightness temperature of the emission is limited to $< 10^{10}$K, and hence requires a source size with a length scale $> 0.5 R_*$ (Equation 1). In order to account for the absence of unpolarized emission for large fractions of the rotational phase of the dwarf, this source would have to be fully occulted by the stellar disk. However, even if such a source region was very close to the surface of the dwarf it would be fully visible for at least $50\%$ and partially visible for at least $75\%$ of the period of rotation of the dwarf. In the case of incoherent gyrosynchrotron emission this would result in a broadly peaked pulse profile present over $75\%$ of the periodic light curve, which does not match the profile of the interpulses detected from 2MASS J0036. We note that this discussion was for the limiting case of emission with brightness temperature of $10^{10}$K. Considering that the degree of circular polarization reaches $75\%$ for the third interpulse (Figure 3), requiring emission at harmonics of the electron cyclotron frequency $\lesssim 20$ \citep{dulk82}, we can further limit the brightness temperature of the emission to be $< 10^{9}$K, which in turn would require a source size with a length scale $> 1.6 R_*$ and a periodic light curve of much lower variation. 

\par We note that both the polarized and unpolarized radio emission from this ultracool dwarf has previously been interpreted as incoherent gyrosynchrotron radiation by \citet{berger05}. However, we correlate Stokes I and V light curves for an observation on 2005 January 11/12, derived from data discussed in that paper, with the light curves for the data obtained on 2006 September 24, and the light curve morphologies are very similar (Figure 3). In particular, the $100\%$ circularly polarized main pulse is once again present, immediately ruling out incoherent gyrosynchrotron emission. Furthermore, we note the presence of a weaker interpulse which also has two sub pulses of right and left circularly polarized emission, similar to what is observed for the third interpulse of the 2006 September 24 light curves. From this discussion we can draw the conclusion that the radio emission, and hence the large-scale magnetic field, is stable over timescales of order 18 months and the emission is not produced by the incoherent gyrosynchrotron mechanism.

\par A more plausible explanation also attributes the unpolarized component of the radio emission detected from 2MASS J0036 to the ECM instability, with the emission being depolarized on traversing the magnetosphere of the dwarf. Considering the interpulse is of similar flux density and duty cycle to the main pulse, it is likely that the emission from both pulses originates in the same extended source region. Assuming ECM emission from a source region located at the magnetic polar regions, beamed in a conical sheet at a large angle to the local magnetic field, we do indeed expect two pulses per period of rotation of the dwarf, occurring when the walls of the conical sheet sweep Earth. Significant depolarization of the ECM emission occurs along one of these intersecting paths in the conical sheet, whereas the emission remains $100\%$ circularly polarized along the other path. Depolarization of coherent emission is a ubiquitous phenomenon observed for coherent bursts detected from the Sun, many of which are attributed to the electron cyclotron maser instability. In fact, any degree of circular polarization is equally probable for solar decimetric and microwave spike bursts, with an average of $25-30\%$ reported by \citet{benz86}. Propagation effects responsible for the depolarization of solar millisecond spike bursts may include scattering or dispersion in the coronal plasma \citep{gudel91, benz97}, strong mode coupling in quasitransverse magnetic field regions \citep[][and references therein]{bastian98} and reflection off boundary layers between regions with large density ratios \citep{melrose06}. 

In considering depolarization of ECM emission from ultracool dwarfs, it is noteworthy that the average degree of circular polarization of the radio emission detected from the L3.5 dwarf, 2MASS J0036, is much higher than that detected from either the M9 dwarf TVLM 513 or the M8.5 dwarf LSR J1835, and therefore the emission from the later type dwarf is subject to much lower depolarization. 2MASS J0036 is a cooler dwarf and should have a predominantly more neutral atmosphere with lower fractional ionization than either TVLM 513 and LSR J1835. This picture is confirmed by the absence of detectable H$\alpha$ and X-ray emission from 2MASS J0036 in contrast to the detection of weak H$\alpha$ emission from both TVLM 513 and LSR J1835, as well as weak X-ray emission from TVLM 513. Therefore, depolarization due to propagation through coronal plasma should be lower for 2MASS J0036 than either TVLM 513 or LSR J1835, consistent with the higher degree of total circular polarization observed for the cooler dwarf.

\section{ROTATION RATES, RADII AND VIEWING GEOMETRY}

\begin{deluxetable*}{lccc}
\tabletypesize{\scriptsize}
\tablecaption{Confirmed Sources of Electron Cyclotron Maser Emission}
\tablewidth{0pt}

\tablehead{
\colhead{ } & \colhead{TVLM 513-46546} & \colhead{2MASS J00361617+1821104} & \colhead{LSR J1835+3259} 
}
\startdata
Spectral Type 				& M9 & L3.5 & M8.5 \\

Distance (pc) 				& 10.6 & 8.8 & 5.7  \\

log($L_{bol}/L_{\odot}$)	& -3.59 & -3.98 & -3.51 \\

Lithium						&	No	&	No &	?	\\

Est. Mass 					& 0.06 - 0.08 & 0.06 - 0.074 & $< 0.083$  \\

Est. Age (Gyr) 				& $> 0.4$  & $>0.8$ & ? \\

$v \sin i$ (km s$^{-1}$) 	& $60$ & 37 & $50 \pm 5$ \\

Rotation Period (hours)		& 1.958 & 3.08 & 2.84 \\

Radius $(R/R_{\odot})$		& $0.097 - 0.109$ & 0.092 - 0.098 & $0.105-0.129$  \\

$i$ ($\arcdeg$)				& 62.5 - 90 & 70 - 90 & $\sim 90$ \\

\enddata
\clearpage

\end{deluxetable*}


\par Based on the radio-derived rotation periods and available $v \sin i$ data for each of the three dwarfs confirmed as sources of ECM emission, LSR J1835, 2MASS J0036 and TVLM 513, we can determine the inclination angle of the rotation axis relative to our line of sight. This analysis is subject to an estimated radius for each dwarf, which in turn is dependent on age and mass. By correlating available data, such as lithium abundance and bolometric luminosity with the evolutionary models put forward by \citet{baraffe98} and \citet{chabrier00}, we attempt to constrain the age, mass and hence radius, for all three dwarfs.

\par TVLM 513 is a young disk M9 dwarf with a bolometric luminosity of log($L_{bol}/L_{\odot}$) = -3.59 \citep{tinney93, tinney95}. Based on this bolometric luminosity and the absence of lithium in its spectrum \citep{reid02}, we can limit the mass of TVLM 513 to $0.06 $M$_{\odot} <$ M$_* < 0.08 $M$_{\odot}$, and infer a minimum age of 400 Myr. Therefore, TVLM 513 is very close to the substellar boundary being either a brown dwarf or an older very low mass star. We estimate a radius in the range $0.097 - 0.109 $ R$_{\odot}$, which together with a period of rotation $p = 1.958$ hours \citep{hallinan07} and a $v \sin i = 60$ km s$^{-1}$ \citep{mohanty03} corresponds to a range of possible inclination angles of $62.5 - 90 \arcdeg$ for the rotation axis of the dwarf.

\par 2MASS J0036 is an L3.5 dwarf with a bolometric luminosity of log($L_{bol}/L_{\odot}$) = -3.98 \citep{reid00, vrba04}. Once again, based on this bolometric luminosity and the absence of lithium from its spectrum, we infer a minimum age of $> 800$ Myr and a mass $0.06 M_{\odot} < M_* < 0.074 M_{\odot}$, the latter pointing to the probable substellar nature of 2MASS J0036. We can therefore estimate a range of possible radii $0.092 - 0.098 $R$_{\odot}$. \citet{schweitzer01} reported a $v \sin i = 15$ km s$^{-1}$ for this dwarf. However, two further separate studies by \citet{jones05} and \citet{zapatero06} yielded values of 38 and 36 km s$^{-1}$ respectively. We use the mean of these latter two values, which are in good agreement with each other, as the $v \sin i$ for 2MASS J0036. Together with a period of rotation of $3.08 \pm 0.05$ hours, this corresponds to a range of inclination angle angles of $70 - 90 \arcdeg$ for the rotation axis of 2MASS J0036 relative to our line of sight. 

\par LSR J1835 is an M8.5 dwarf with a bolometric luminosity of log($L_{bol}/L_{\odot}$) = -3.51 \citep{reid03}. No confirmation is currently available in the literature as to whether the lithium absorption feature is present or absent in the spectrum of LSR J1835, and therefore we cannot use this property, together with the bolometric luminosity, to constrain the mass, age and hence radius of the dwarf. However, based on the period of rotation of 2.84 hours for LSR J1835 derived from the radio data and a $v \sin i = 50 \pm 5$ km s$^{-1}$ \citep{berger07b}, it becomes apparent that even for the limiting case of a viewing angle of $90 \arcdeg$, a radius $ \geq 0.117 \pm 0.012$ R$_{\odot}$ is required. We consider three possibilities to account for such a large radius. 1) LSR J1835 is of young or intermediate age and, due to  bolometric luminosity constraints, is probably a brown dwarf. 2) The $v \sin i$ of the dwarf is close to the lower limit of the error bars, yielding a radius $ \sim 0.105$ R$_{\odot}$. 3) LSR J1835 is an evolved very low mass star with a larger radius than predicted by current theoretical models. A discrepancy of $\sim 10\%$ between the theoretically predicted radii of low mass stars and those directly measured, either through observations of eclipsing binary systems or interferometric studies of single stars, has been reported \citep[][and references therein]{ribas06}, possibly due to the effects of activity \citep{mullan01, lopez05, lopez07, morales08} or metallicity \citep{bergerb06}. A recent study by \citet{lopez07} has shown that this discrepancy is reduced for stars with mass below $0.35$ M$_{\odot}$ and these fully convective stars have radii closer to the expected theoretical predictions. However, this study was limited to relatively slow rotators for the fully convective sample and no ultracool dwarfs were included. We note that LSR J1835+3259 is a rapidly rotating, active ultracool dwarf with tentative evidence for a larger radius than expected from theoretical models. Further high resolution spectroscopic observations of LSR J1835 are essential to establish whether lithium is present in the spectrum of LSR J1835, thereby constraining the mass and age of this ultracool dwarf, as well as further $v \sin i$ measurements to refine the radius estimate. Building on the radius limitations discussed here, this will present a unique opportunity to investigate if the radius of a rapidly rotating, active ultracool dwarf is in line with those predicted by theoretical models.   

\par TVLM 513, 2MASS J0036 and LSR J1835 all have a rotation axis which is close to perpendicular to our line of sight (Table 1). The chance probability of all 3 dwarfs having inclination angles $> 62.5 \arcdeg$, which we note as the lower limit of the inclination angle for TVLM 513, is $< 3 \times 10^{-2}$.  This is particularly significant in the context of the tentative relationship observed between $v \sin i$ and radio activity for ultracool dwarfs, whereby all but one of the dwarfs detected thus far at radio frequencies have $v \sin i > 25$ km s$^{-1}$ \citep{berger07b} . \citet{berger07b} suggest this relationship may underly a dependence on rapid rotation for the detection of emission, and indeed the rotation velocities confirmed for TVLM 513, LSR J1835 and 2MASS J0036 support this hypothesis. However, an intriguing alternative possibility is that the relationship observed between $v \sin i$ and radio activity is not due to a dependence on rotation velocity $v$, but rather indicates a dependence on inclination angle $i$. Considering that radio emission from ultracool dwarfs is attributed to the ECM instability, a mechanism that produces highly directive emission, the detection of targets with higher $v \sin i$ may underly a geometrical selection effect. For illustrative purposes, we consider the specific case of a large-scale dipolar magnetic field tilted at an angle $\theta$ relative to the rotation axis of the dwarf, which itself has an inclination angle $i$ relative to our line of sight. If we assume ECM emission is generated at the poles of the large-scale magnetic field and beamed perpendicular to the field in the source region, emission is detected over the range of rotational phase when this large-scale magnetic field lies approximately in the plane of the sky. Without factoring in the thickness of the wall of the emission cone and the extent of the source region, we can approximate that if $\theta + i < 90 \arcdeg$, then the large-scale field will never lie in the plane of the sky, and the ECM emission will never be detected. We note that the depolarization of the ECM emission may play a role in reducing the directivity of the emission. \citet{donati06} have shown that the large-scale dipolar field of the rapidly rotating convective star, V374 Peg has a large-scale dipolar component that is close to aligned to the rotation axis. If the large-scale magnetic fields of ultracool dwarfs are also close to aligned, with very low $\theta$, then only a small fraction would be detectable sources of ECM emission, with the detected sources having large values of inclination angle $i$. This is certainly consistent with the large inclination angles confirmed for TVLM 513, 2MASS J0036 and LSR J1835. Direct evidence for the importance of geometry for the detection of emission can be derived from the light curve of 2MASS J0036 (Figure 2), where the emission periodically drops below the detection limit, due to the unfavorable beaming of the emission at certain phases of rotation of the dwarf. Finally, we also note that the separation in phase of two pulses of ECM emission from the same source region is also dependent on $\theta$ and $i$. Considering again the example of emission beamed perpendicular to a large-scale dipolar magnetic field, we can define the separation in phase between the two pulses as $\alpha = \arccos(\cot \theta \cot i)/180\arcdeg$.
 
\section{SUMMARY AND CONCLUSIONS}

\begin{deluxetable*}{ccccc}
\tabletypesize{\scriptsize}
\tablecaption{Sample of M dwarfs and Ultracool dwarfs detected at 8.44 GHz}
\tablewidth{0pt}

\tablehead{
\colhead{Name} & \colhead{Spectral Type} & \colhead{8.5 GHz Flux (mJy)} & \colhead{Distance (pc)} &\colhead{Radio Luminosity (ergs s$^{-1}$ Hz$^{-1}$}
}
\startdata

YY Gem AB 				& M0.5 & 0.268 & 14.95 & $7.16 \times 10^{13}$ \\
BD+13 2618 				& M0.5 & 0.115 & 11.43 & $1.80 \times 10^{13}$ \\
AU Mic					& M1 & 1.000 & 9.94 & $1.18 \times 10^{14}$ \\
FK Aqr					& M1.5 & 0.360 & 8.68 & $3.24 \times 10^{1})$ \\
V1285 Aql				& M3 & 0.087 & 11.59 & $1.40 \times 10^{13}$ \\	
NSV 14168				& M3 & 0.082 & 4.01 & $1.58 \times 10^{13}$ \\
V998 Ori				& M3.5 & 0.1.63 & 12.77 & $3.18 \times 10^{13}$ \\	
GJ 644 AB				& M3.5 & 0.475 & 6.50 & $2.40 \times 10^{13}$ \\
V1216 Sgr				& M3.5 & 0.165 & 2.97 & $1.74 \times 10^{12}$ \\
FL Aqr					& M3.5 & 0.245 & 8.68 & $2.21 \times 10^{13}$ \\
EQ Peg					& M3.5 & 0.459 & 6.25 & $2.14 \times 10^{13}$ \\
DO Cep					& M4 & 0.082 & 4.01 & $1.58 \times 10^{12}$ \\
LHS 25					& M4.5 & 0.311 & 5.04 & $9.47 \times 10^{12}$ \\
YZ CMi					& M4.5 & 0.323 & 5.92 & $1.36 \times 10^{13}$ \\
AD Leo					& M4.5 & 0.410 & 4.69 & $1.08 \times 10^13$ \\
AT Mic AB				& M4.5 & 1.985 & 10.22 & $2.48 \times 10^{14}$ \\
LHS 3966				& M4.5 & 0.226 & 6.25 & $1.06 \times 10^{13}$ \\
UV Ceti AB				& M5.5 & 1.650 & 2.62 & $1.36 \times 10^{13}$ \\
LHS 1070ABC				& M5.5 & 0.161 & 7.39 & $1.05 \times 10^{13}$ \\
LHS 3003				& M7 & 0.270 & 6.34 & $1.30 \times 10^{13}$ \\
LP 349-25AB				& M8 & 0.365 & 14.79 & $9.56 \times 10^{13}$ \\
DENIS 1048-3956			& M8 & 0.149 & 4.04 & $2.73 \times 10^{12}$ \\
LSR J1835+3259			& M8.5 & 0.720 & 5.67 & $2.77 \times 10^{13}$ \\
TVLM 513-46546			& M9 & 0.430 & 10.59 & $5.77 \times 10^{13}$ \\
LP 944-20				& M9 & 0.074 & 4.97 & $2.18 \times 10^12$ \\
BRI 0021-0214			& M9.5 & 0.083 & 11.55 & $1.32 \times 10^13$ \\
2M 0523-14				& L2.5 & 0.231 & 13.4 & $4.97 \times 10^13$ \\
2M 0036+18				& L3.5 & 0.134 & 8.76 & $1.23 \times 10^13$ \\

\enddata
\tablecomments{M dwarf radio luminosities : \citep{gudel93b}. Ultracool dwarf radio luminosities: \citep{berger01, berger02, berger05, berger06, berger07a, berger07b, burgasser05, osten06, phan-bao07, hallinan06, hallinan07, antonova07}}

\end{deluxetable*}

\begin{figure}
\plotone{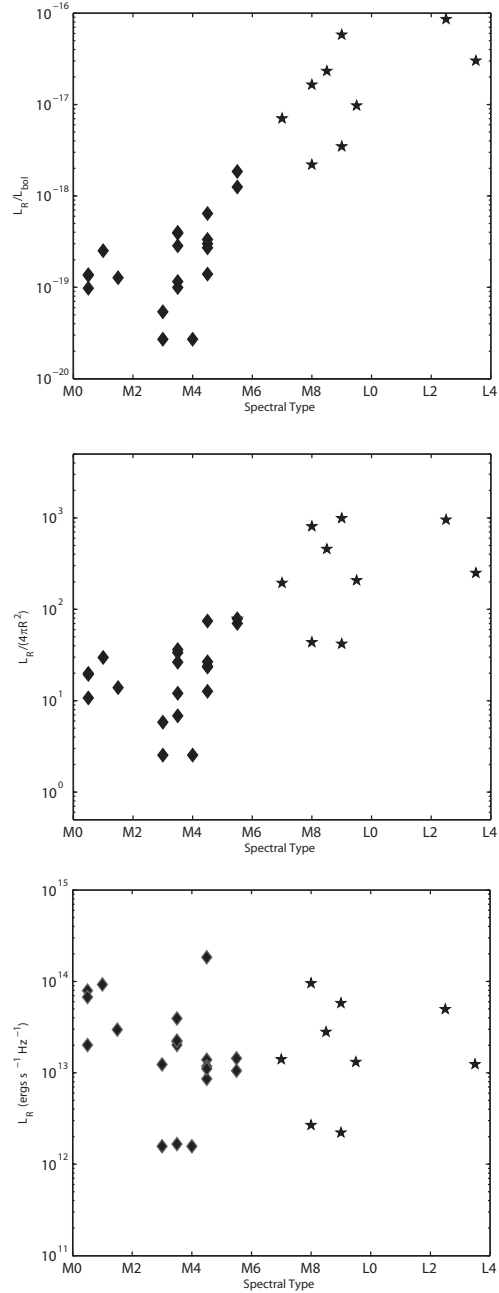}
\caption{Radio luminosities are maintained from early M dwarfs through to late M and L dwarfs and are largely independent of  bolometric luminosity and the surface area of the stellar disk. [Top] Radio to bolometric luminosity ratio vs spectral type for M dwarfs (diamonds) and ultracool dwarfs (stars). [Middle] Radio luminosity normalized with the surface flux vs spectral type for M dwarfs (diamonds) and ultracool dwarfs (stars). [Bottom] Radio luminosity vs spectral type for M dwarfs (diamonds) and ultracool dwarfs (stars). M dwarf radio luminosities and ultracool dwarf radio luminosities are calculated for detections at 8.44 GHz with the data summarized in Table 2.}
\end{figure}

\par We have reported on radio observations of two ultracool dwarfs, the M8.5 dwarf LSR J1835 and the L3.5 dwarf 2MASS J0036, that confirm the ECM instability as the dominant mechanism producing radio emission in the magnetospheres of ultracool dwarfs. Both periodic ($p = 2.84$ hours), $100\%$ circularly polarized, narrow duty cycle pulses and quiescent unpolarized emission are detected from LSR J1835 with the periodic pulses being of similar nature to those previously detected from TVLM 513 \citep{hallinan07}. 2MASS J0036, on the other hand, is a source of periodic pulses, both $100\%$ circularly polarized and unpolarized, with a much broader morphology, with no evidence for a quiescent, aperiodic component to the radio emission. Crucially, brightness temperature limitations imposed on the periodic, unpolarized radio emission detected from this dwarf rule out incoherent emission mechanisms such as gyrosynchrotron radiation. The unpolarized emission from this and other ultracool dwarfs is therefore attributed to depolarized ECM emission, with depolarization of the ECM emission possibly occurring due to propagations effects such as scattering. The confirmation of ECM emission from both dwarfs requires large-scale stable magnetic field configurations with field strengths at the magnetic poles $\geq 3$ kG and $\geq 1.7$ kG for LSR J1835 and 2MASS J0036 respectively. We note that, in the case of 2MASS J0036, this represents the first confirmation of kG magnetic fields for L dwarfs establishing strong magnetic dynamo action out to spectral type L3.5. 

\par ECM emission from the polar regions of a large-scale, high strength magnetic field is consistent with the kilogauss field strengths confirmed for ultracool dwarfs by \citet{reiners07} as well as the large-scale, stable, magnetic fields predicted by current dynamo models and confirmed by observation for fully convective low mass objects \citep{dobler06, chabrier06, donati06, browning07}. The long-term stability in the morphology of the periodic light curve of the radio emission from 2MASS J0036 also establishes the large-scale magnetic field of this dwarf to be stable over time scales $> 18$ months. A high degree of long term variability has been previously confirmed for the radio emission from some ultracool dwarfs, including the emission from both TVLM 513 and LSR J1835 \citep{osten06, antonova07, hallinan06, hallinan07, berger07a, berger07b}. In the case of TVLM 513, it has been suggested that this long term variability may be associated with a change in field configuration on timescales $\lesssim 1$ yr \citep{berger07b}. However, this is inconsistent with the long-term stability over time scales $> 1$ yr confirmed for the large-scale magnetic field of the rapidly rotating, fully convective dwarf star, V374 Peg \citep{donati06, morales08}. Rather, we attribute long term variability in the levels of radio emission from ultracool dwarfs to variation in the local plasma conditions at the magnetic poles of a stable large-scale magnetic field, possibly triggered by magnetic reconnection events such as those reported by \citet{stelzer06b} and \citet{schmidt07}. If this is indeed the case, long term radio monitoring of ultracool dwarfs should reveal periodic structures that vary, disappear and possibly reappear due to long term variations in the plasma conditions in the associated compact regions. Such structures should be confined to a particular range of phase of rotation of the dwarf governed by the topology of the large-scale stable magnetic field. 

\par The onset of ECM emission is also consistent with the observed H$\alpha$ and X-ray activity from ultracool dwarfs. The sharp drop observed in emission associated with the non-radiative heating of plasma to chromospheric and coronal plasma indicates a trend towards cooler and more neutral magnetospheric conditions with later spectral type, with a global drop occurring in the fractional ionization of magnetospheric plasma. However, the detection of strong flares in these wave bands is in line with the confirmation that high strength magnetic fields are maintained across the same spectral range. Therefore conditions occur where the electron cyclotron frequency dominates the plasma frequency for large portions of the ultracool dwarf magnetosphere, favoring the generation of ECM emission. 

\par Based on available $v \sin i$ data in the literature and rotation periods derived from the periodic radio data for TVLM 513, LSR J1835 and 2MASS J0036, we determine that the rotation axes of all three dwarfs are approximately perpendicular to our line of sight. In light of the inherent directivity of coherent ECM emission, this may imply a geometrical selection effect for the radio emission from ultracool dwarfs, whereby if the large scale magnetic field of the ultracool dwarf is close to aligned (low $\theta$) and the inclination angle $i$ of the rotation axis relative to our line of sight is also small such that $\theta + i < 90 \arcdeg$, then ECM emission will not be detected. If such a geometrical selection effect is applicable, it would account for the relationship between radio activity and $v \sin i$ observed for ultracool dwarfs. We also determine the radius of the dwarf LSR J1835+3259 to be $\geq 0.117 \pm 0.012$ R$_{\odot}$. The implied size of the radius, together with its bolometric luminosity, implies a substellar classification for this ultracool dwarf, or alternatively, that current theoretical models underestimate the radii of ultracool dwarfs.

\par The observed trends in radio luminosity from early and mid M type dwarfs through to ultracool dwarfs have been discussed recently by \citet{berger06} and \citet{audard07}. \citet{berger06} noted that radio to bolometric luminosity ratio, $L_{R}/L_{bol}$, increases with later spectral type from early and mid M dwarfs through to ultracool dwarfs (Figure 4). However, as pointed out by \citet{audard07}, this trend in $L_{R}/L_{bol}$ can be attributed to the reduction in effective temperature and hence bolometric luminosity across this spectral range, and suggest an alternative measure of radio activity, the surface flux $L_{R}/4\pi R^2$. In terms of surface flux, radio luminosity is seen to increase from early to mid M dwarfs and reach a plateau for ultracool dwarfs. However, once again, the observed trend can be attributed to the decrease in radius with $T_{eff}$, a point duly noted by \citet{audard07}. In fact, if considered independently of bolometric luminosity and radii, radio luminosities are steadily maintained from early type M dwarfs through to late M and early L type dwarfs (Figure 4). This is extremely significant when considering that bolometric luminosity drops by three orders of magnitude over this spectral range. Similarly, brightness temperature for the quiescent radio emission from active M dwarfs, generally attributed to gyrosynchrotron radiation, is usually calculated by assuming a source size of the order of size of the stellar disk, or a multiple of this value. If this was indeed the case, we would expect a sharp drop in luminosity from early M to late M and L type dwarfs. The fact that luminosities are maintained from early M dwarfs through to late M and L dwarfs, points to a common emission mechanism that is sustained at the lower end of the main sequence and is independent of bolometric luminosity and the surface area of the stellar disk. 

\par In light of the confirmation of the ECM instability as a source of both circularly polarized and unpolarized radio emission from ultracool dwarfs, this mechanism must also now be reconsidered as a contributing source of quiescent emission from M dwarfs, active binaries and protostars. In particular, we note that the luminosity of ECM emission produced in compact regions at the magnetic poles should be largely independent of bolometric luminosity and surface area of the stellar disk. If the ECM instability proves to be a viable source of quiescent unpolarized radio emission, indistinguishable in temporal and polarization characteristics from gyrosynchrotron radiation, it will have important ramifications for the magnetic field diagnostics derived from radio observations of stellar sources.



\acknowledgments

GH and AG gratefully acknowledge the support of Science Foundation Ireland (Grant Number 07/RFP/PHYF553). Armagh Observatory is grant-aided by the N. Ireland Dept. of Culture, Arts \& Leisure. We thank the referee, Manuel G\"udel, for his careful reading of the manuscript and his helpful suggestions for improving the manuscript.






\begin{thebibliography}{}

\bibitem[Antonova et al.\ (2007)]{antonova07} Antonova, A., Doyle, J. G., Hallinan, G., Golden, A., \& Koen, C. 2007, A\&A, 472, 257
\bibitem[Audard et al.\ (2007)]{audard07} Audard, M., Osten, R. A., Brown, A., Briggs, K. R., G\"udel, M., Hodges-Kluck, E., \& Gizis, J. E. 2007, A\&A, 471, L63
\bibitem[Baraffe et al(1998)]{baraffe98} Baraffe, I., Chabrier, G., Allard, F., \& Hauschildt, P. H. 1998, A\&A, 337, 403
\bibitem[Bastian et al.\ (1998)]{bastian98} Bastian, T. S., Benz, A. O., \& Gary, D. E. 1998, ARA\&A 36, 131
\bibitem[Benz (1986)]{benz86} Benz, A. O. 1986, Sol. Phys., 104, 99
\bibitem[Benz \& G\"udel (1994)]{benz94} Benz, A. O., \& G\"udel, M. 1994, A\&A, 285, 621
\bibitem[Benz \& Pianezzi (1997)]{benz97} Benz, A. O. \& Pianezzi, P. 1997, A\&A, 323, 250
\bibitem[Berger (2002)]{berger02} Berger, E. 2002, \apj, 572, 503
\bibitem[Berger et al.\ (2001)]{berger01} Berger, E., et al. 2001, Nature, 410, 338
\bibitem[Berger et al.\ (2005)]{berger05} Berger, E., et al. 2005, \apj, 627, 960
\bibitem[Berger (2006)]{berger06} Berger, E. 2006, \apj, 648, 629
\bibitem[Berger et al.\ (2007a)]{berger07a} Berger, E., et al. 2007a, \apj submitted (astro-ph/0710.3383) \\
\bibitem[Berger et al.\ (2007b)]{berger07b} Berger, E., et al. 2007b, \apj submitted (astro-ph/0708.1511) \\
\bibitem[Berger et al.\ (2006)]{bergerb06} Berger, D. H., et al. 2006, \apj, 644, 475
\bibitem[Bingham et al.\ (2001)]{bingham01} Bingham, R., Cairns, R. A., \& Kellett, B. J. 2001, A\&A, 370, 1000
\bibitem[Browning (2007)]{browning07} Browning 2007, \apj submitted (astro-ph/0712.1603) \\
\bibitem[Burgasser et al.(2000)]{burgasser00} Burgasser, A. J., Kirkpatrick, J. D., Reid, I. N., Liebert, J., Gizis, J. E., \& Brown, M. E. 2000, \aj, 120, 473
\bibitem[Burgasser \& Putman (2005)]{burgasser05} Burgasser, A. J., \& Putman, M. E. 2005, \apj, 626, 486
\bibitem[Chabrier et al.(2000)]{chabrier00}Chabrier, G., Baraffe, I., Allard, F., \& Hauschildt, P. 2000, ApJ, 542, 464
\bibitem[Chabrier \& K\"uker (2006)]{chabrier06} Chabrier, G., \& K\"uker, M. 2006, A\&A, 446, 1027
\bibitem[Close et al.\ (2003)]{close03} Close, L. M., Siegler, N., Freed, M., \& Biller, B. 2003, \apj, 587, 407
\bibitem[Dobler et al.\ (2006)]{dobler06} Dobler, W., Stix, M., \& Brandenburg, A. 2006, \apj, 638, 336
\bibitem[Donati et al.\ (2006)]{donati06} Donati, J.-F., Forveille, T., Cameron, A. C., Barnes, J. R., Delfosse, X., Jardine, M. M., Valenti, J. A. 2006, Science, 311, 633
\bibitem[Donati et al.\ (2007)]{donati07} Donati, J.-F.et al. 2007, Donati J.-F. et al., 2007, Proceedings of the 14th Meeting on Cool Stars, Stellar Systems and the Sun, in press (astro-ph/0702159) 
\bibitem[Dulk (1985)]{dulk85} Dulk, G. A. 1985, ARA\&A, 23, 169
\bibitem[Dulk \& Marsh (1982)]{dulk82} Dulk, G. A., \& Marsh, K. A. 1982, \apj, 259, 350
\bibitem[Ergun et al.\ (2000)]{ergun00} Ergun, R. E., Carlson, C. W., McFadden, J. P., Delory, G. T., Strangeway, R. J., \& Pritchett, P. L. 2000, ApJ, 538, 456
\bibitem[Fleming et al.(2003)]{fleming03} Fleming, T. A., Giampapa, M. S., \& Garza, D. 2003, \apj, 594, 982
\bibitem[Fuhrmeister \& Schmitt (2004)]{fuhrmeister04} Fuhrmeister, B., \& Schmitt, J. H. M. M. 2004, A\&A, 420, 1079
\bibitem[Gizis et al.\ (2000)]{gizis00} Gizis, J. E., Monet, D. G., Reid, I. N., Kirkpatrick, J. D., Liebert, J., \& Williams, R. J. 2000, \aj, 120, 1085
\bibitem[G\"udel (2002)]{gudel02} G\"udel, M. 2002, ARA\&A, 40, 217
\bibitem[G\"udel \& Benz (1993)]{gudel93} G\"udel, M., \& Benz, A. O. 1993, \apj, 405, L63
\bibitem[G\"udel et al.\ (1993)]{gudel93b} G\"udel, M., Schmitt, J. H. M.M., Bookbinder, J. A., \& Fleming, T. A. 1993, ApJ, 415, 236
\bibitem[G\"udel \& Zlobec (1991)]{gudel91}G\"udel, M. \& Zlobec, P. 1991, A\&A, 245, 299
\bibitem[Hallinan et al.\ (2006)]{hallinan06} Hallinan, G., Antonova, A, Doyle, J. G., Bourke, S., Brisken, W. F. \& Golden, A. 2006, \apj, 653, 690
\bibitem[Hallinan et al.\ (2007)]{hallinan07} Hallinan, G., et al. 2007, ApJ, 663, L25
\bibitem[Jones et al.(2005)]{jones05} Jones, H. R. A., Pavlenko, Ya., Viti, S., Barber, R. J., Yakovina, L. A., Pinfield, D., \& Tennyson, J. 2005, MNRAS, 358, 105
\bibitem[Kellett et al.\ (2002)]{kellett02} Kellett, B. J., Bingham, R., Cairns, R. A., \& Tsidouki, V. 2002, MNRAS, 329, 102
\bibitem[Kirkpatrick et al.\ (1997)]{kirkpatrick97} Kirpatrick, J. D., Henry, T. J., \& Irwin M. J. 1997, \aj, 113, 1421
\bibitem[Lane et al.\ (2007)]{lane07} Lane, C., et al. 2007, ApJ, 668, L25
\bibitem[Liebert et al.\ (2003)]{liebert03} Liebert, J., Kirkpatrick, J. D., Cruz, K. L., Reid, I. N., Burgasser, A. J., Tinney, C. G., \& Gizis, J. E. 2003, \aj, 125, 343
\bibitem[L\'opez-Morales \& Ribas (2005)]{lopez05} L\'opez-Morales, M. \& Ribas, J. C. 2005, \apj, 631, 1120
\bibitem[L\'opez-Morales (2007)]{lopez07} L\'opez-Morales,M. 2007, \apj, 660, 732
\bibitem[Meyer \& Meyer-Hofmeister (1999)]{meyer99} Meyer, F., \& Meyer-Hofmeister, E. 1999, A\&A, 341, L23
\bibitem[Mohanty et al.\ (2002)]{mohanty02} Mohanty, S., Basri, G., Shu, F., Allard, F., \& Chabrier, G. 2002, \apj, 571, 469
\bibitem[Mohanty \& Basri (2003)]{mohanty03} Mohanty, S., \& Basri, G. 2003, \apj, 583, 451
\bibitem[Morales et al.\ (2008)]{morales08} Morales, J. C., \& Ribas, I., \& Jordi, C. 2008, A\&A, 478, 507
\bibitem[Melrose \& Dulk (1982)]{melrose82} Melrose, D. B., \& Dulk, G. A. 1982, \apj, 259, 844
\bibitem[Melrose (2006)]{melrose06} Melrose, D. B. 2006, \apj 637, 1113
\bibitem[Morin et al.\ (2007)]{morin07} Morin, J., et al. 2007, MNRAS submitted astro-ph/(0711.1418)
\bibitem[Mullan \& McDonald (2001)]{mullan01} Mullan, D. J. \& MacDonald, J. 2001, \apj, 559, 353
\bibitem[Neuh\"aser et al.(1999)]{neuhauser99} Neuha\"user, R., et al. 1999, A\&A, 343, 883
\bibitem[Osten et al.\ (2006)]{osten06} Osten, R. A., Hawley, S. L., Bastian, T. S., \& Reid, I. N. 2006, ApJ, 637, 518
\bibitem[Osten \& Bastian(2006)]{ostenbastian06} Osten, R. A., \& Bastian, T. S. 2006, ApJ, 637, 1016
\bibitem[Phan-Bao et al.\ (2007)]{phan-bao07} Phan-Bao, N., Osten, R. A., Lim, J., Mart\'in, E. L. \& Ho, P. T. P. 2007, \apj, 658, 553 
\bibitem[Reid et al.\ (1999)]{reid99} Reid, I. N., Kirkpatrick, J. D., Gizis, J. E., \& Liebert, J. 1999, \apj, 527, L105
\bibitem[Reid et al.\ (2000)]{reid00} Reid, I. N., Kirkpatrick, J. D., Gizis, J. E., Dahn, C. C., Monet, D. G., Williams, R. J., Liebert, J., \& Burgasser, A. J. 2000, AJ, 119, 369
\bibitem[Reid et al.(2002)]{reid02} Reid, I. N.,Kirkpatrick, J. D., Liebert, J., Gizis, J. E., Dahn, C. C.,\& Monet,D.G. 2002, \aj, 124, 519
\bibitem[Reid et al.\ (2003)]{reid03} Reid, I. N., et al. 2003, AJ, 125, 354
\bibitem[Reiners \& Basri (2007)]{reiners07} Reiners, A., \& Basri, G. 2007, \apj, 656, 1121
\bibitem[Ribas (2006)]{ribas06} Ribas, I. 2006, Ap\&SS, 304, 89
\bibitem[Rockenfeller et al.(2006)]{rockenfeller06} Rockenfeller, B., Bailer-Jones, C. A. L., Mundt, R., Ibrahimov, M. A. 2006 MNRAS, 367, 407
\bibitem[Rutledge et al.(2000)]{rutledge00} Rutledge, R. E., Basri, G., Mart\`in, E. L., \& Bildsten, L. 2000, \apj, 538, L141
\bibitem[Schmidt et al.\ (2007)]{schmidt07} Schmidt, S. J., Cruz, K. L., Bongiorno, B. J., Liebert, J., \& Reid, I. N. 2007, \aj, 133, 2258
\bibitem[Schweitzer et al.(2001)]{schweitzer01} Schweitzer, A., Gizis, J. E., Hauschildt, P. H., Allard, F., \& Reid, I. N. 2001, ApJ, 555, 368
\bibitem[Stelzer (2004)]{stelzer04} Stelzer, B. 2004, \apj, 615, L153
\bibitem[Stelzer et al.\ (2006a)]{stelzer06a} Stelzer, B., Micela, G., Flaccomio, E., Neuh\"auser, R., \& Jayawardhana, R. 2006a, A\&A, 448, 293
\bibitem[Stelzer et al.\ (2006b)]{stelzer06b}Stelzer, B., Schmitt, J. H. M. M., Micela, G., \& Liefke, C. 2006b, A\&A, 460, L35
\bibitem[Tinney et al.(1993)]{tinney93} Tinney, C. G., Mould, J. R., \& Reid, I. N. 1993, AJ, 105, 1045
\bibitem[Tinney et al.(1995)]{tinney95} Tinney, C. G., Reid, I. N.,Gizis, J., \& Mould, J. R. 1995, \aj, 110, 3014
\bibitem[Treumann (2006)]{treumann06} Treumann, R. A. 2006, A\&A Rev. 13, 229
\bibitem[Vrba et al.\ (2004)]{vrba04} Vrba, F. J., et al. 2004, \aj, 127, 2948
\bibitem[West et al.\ (2004)]{west04} West, A. A., et al. 2004, \aj, 128, 426
\bibitem[White \& Franciosini (1995)]{white95} White, S. M., \& Franciosini, E. 1995, ApJ, 444, 342
\bibitem[Zapatero Osorio et al.\ (2006)]{zapatero06} Zapatero Osorio, M. R., Mart\'in, E. L., Bouy, H., Tata, R., Deshpande, R. \& Wainscoat, R. 2006, \apj, 647, 1405
\bibitem[Zarka (1998)]{zarka98} Zarka, P. 1998, J. Geophys. Res., 103, 20159


\end{thebibliography}

\end{document}